\documentclass[jcp,aip,10pt,floatfix,twocolumn,showpacs]{revtex4-1} 
\usepackage{blindtext}
\usepackage{color,soul}
\usepackage{graphicx}
\graphicspath{Image/Image/}
\usepackage{amsmath, amsthm, amssymb}
\usepackage{amsthm}
\usepackage{multirow}
\usepackage{makecell}
\usepackage[table]{xcolor}
\definecolor{Gray}{gray}{0.85}
\definecolor{LightCyan}{rgb}{0.88, 1, 1}
\definecolor{Apricot}{rgb}{0.98, 0.81, 0.69}
\usepackage[normalem]{ulem}
\usepackage{soul}
\newcommand{\be}{\begin{equation}}
	\newcommand{\ee}{\end{equation}}
\newcommand{\bea}{\begin{eqnarray}}
	\newcommand{\eea}{\end{eqnarray}}
\newcommand{\beginsupplement}{
	\setcounter{table}{0}
	\renewcommand{\thetable}{S\arabic{table}}
	\setcounter{figure}{0}
	\renewcommand{\thefigure}{S\arabic{figure}}
}
\usepackage{color,soul}
\usepackage{graphicx}
\usepackage{siunitx}
\usepackage[breaklinks=true,colorlinks=true,linkcolor=blue,urlcolor=blue,citecolor=blue]{hyperref}

\usepackage{amsmath, amsthm, amssymb}
\usepackage{amsthm}
\usepackage{multirow}
\usepackage[normalem]{ulem} 
\usepackage{soul}
\usepackage{makecell}
\usepackage[table]{xcolor}
\definecolor{Gray}{gray}{0.85}
\definecolor{LightCyan}{rgb}{0.88, 1, 1}
\definecolor{Apricot}{rgb}{0.98, 0.81, 0.69}
\usepackage{threeparttable}

\usepackage[rightcaption,]{sidecap}
\sidecaptionvpos{figure}{c}
\usepackage[utf8]{inputenc}
\usepackage{amsmath}
\usepackage{amssymb}
\usepackage{graphicx}
\usepackage{hyperref}
\usepackage[export]{adjustbox}
\usepackage[update]{epstopdf}
\usepackage{multirow}
\usepackage{placeins}

\begin{document}
\title{The conformational phase diagram of neutral polymers in the presence of attractive crowders}
\author{Hitesh Garg}
\email{hiteshgarg@imsc.res.in}
\affiliation{The Institute of Mathematical Sciences, C.I.T. Campus,
Taramani, Chennai 600113, India}
\affiliation{Homi Bhabha National Institute, Training School Complex, Anushakti Nagar, Mumbai, 400094, India}

\author{R Rajesh}
\email{rrajesh@imsc.res.in}
\affiliation{The Institute of Mathematical Sciences, C.I.T. Campus,
Taramani, Chennai 600113, India}
\affiliation{Homi Bhabha National Institute, Training School Complex, Anushakti Nagar, Mumbai, 400094, India}

\author{Satyavani Vemparala}
\email{vani@imsc.res.in}
\affiliation{The Institute of Mathematical Sciences, C.I.T. Campus,
Taramani, Chennai 600113, India}
\affiliation{Homi Bhabha National Institute, Training School Complex, Anushakti Nagar, Mumbai, 400094, India}

\date{\today}

\begin{abstract}
Extensive coarse grained molecular dynamics simulations are performed to investigate the conformational phase diagram of a neutral polymer in the presence of attractive crowders. We show that, for low crowder densities, the polymer predominantly shows three phases as a function of both intra polymer and polymer-crowder interactions: (1) weak intra polymer and weak polymer-crowder attractive interactions induce extended or coil polymer conformations (phase E) (2) strong intra polymer and relatively weak polymer-crowder attractive interactions induce collapsed or globular conformations (phase CI) and (3) strong polymer-crowder attractive interactions, regardless of intra polymer interactions, induce a second collapsed or globular conformation that encloses bridging crowders (phase CB). The detailed phase diagram is obtained by determining the phase boundaries delineating the different phases based on an analysis of the radius of gyration as well as bridging crowders.  The dependence of the phase diagram on strength of crowder-crowder attractive interactions and crowder density is clarified. We also show that when the crowder density is increased, a third collapsed phase of the polymer emerges for weak intra polymer attractive interactions. This crowder density induced compaction is shown to be enhanced by stronger crowder-crowder attraction and is different from the depletion induced collapse mechanism which is primarily driven by repulsive interactions. We also provide a unified explanation of the observed reentrant swollen/extended conformations of earlier simulations of weak and strongly self interacting polymers in terms of crowder-crowder attractive interactions.
\end{abstract}

\maketitle
\section{Introduction}
Polymers are known to undergo extended to collapsed (or coil to globule) transition under a variety of conditions. The simplest of these conditions is the nature of the solvent: good solvents promote extended structures of polymers to maximize the polymer-solvent interactions while poor solvents induce collapsed structures to maximize intra polymer interactions~\cite{de1979scaling,de1990polymers}. Polymers can also undergo collapse transition as the temperature is lowered or increased, depending on whether polymers exhibit an upper critical solution temperature (UCST)~\cite{sun1980coil,niskanen2017manipulate} or lower critical solution temperature, (LCST)~\cite{schild1990microcalorimetric,maeda2002change} behavior. Further, mixtures of solvents can have significant and often counter intuitive effects which are exemplified in phenomena of cosolvency and cononsolvency exhibited by mixed solvents~\cite{masegosa1984preferential, tanaka2008temperature,mukherji2017depleted}. Polymers can also collapse due to entropic effects. Even in a system with purely repulsive interactions, an increase in solvent density can lead to a collapsed phase due to depletion induced attraction between the monomers~\cite{mukherji2017depleted,lekkerkerker2011depletion}. Charged polymers have also been shown to undergo counterintuitive extended to collapse transition as the polymer charge density is increased, in the presence of oppositely charged counterions, regardless of solvent conditions~\cite{liu2003polyelectrolyte,tom2016mechanism,winkler1998collapse,jeon2007necklace,varghese2011phase}.

For neutral polymers, while the polymer configurations are determined largely by the nature of the solvent, it has been shown that the presence of crowder molecules can complicate this simple picture~\cite{nayar2020small,zangi2009urea,mardoum2018crowding,nakano2017model,shin2015kinetics,liu2020non,taylor2020effects}. Understanding the role of crowders on the conformation of polymers is of particular importance as biological polymers are often found in crowded environments filled with crowders of different sizes and shapes~\cite{zimmerman1991estimation,ellis2003join,ellis2001macromolecular,zimmerman1993macromolecular,zhou2008macromolecular,rivas2004life}.  Earlier studies~\cite{zhou2008macromolecular,asakura1958interaction,bhat1992steric,kang2015effects} have shown that the onset of collapsed conformations of neutral polymers with repulsive interactions in the presence of high density of crowders is a function of the size of the crowder particles.  

Counter to the simple view that strong attractive interaction between polymer and the solvent/crowder molecules (mimicking good solvent conditions) extends the polymer, recent studies have observed a counterintuitive collapse transition driven by increasing polymer-crowder attraction~\cite{antypov2008computer,heyda2013rationalizing,rodriguez2015mechanism,sagle2009investigating,huang2021chain,ryu2021bridging,brackley2020polymer,brackley2013nonspecific,barbieri2012complexity}.  Crowder particles have been suggested to act as bridges/glue between distant monomers along the polymer backbone,  inducing collapse via  increased effective intra polymer attractive interactions even in polymers with repulsive bare intra polymer interactions. An experimental realization of such bridging induced collapse can be seen in poly(N-isopropylacrylamide) PNIPAM polymers. In the presence of urea molecules, these polymers have been shown to undergo collapse transition, possibly due to the direct interaction of urea with the polymer via hydrogen bonds, inducing effective short range intra polymer attraction~\cite{sagle2009investigating}. 

The conformation of a neutral polymer is primarily determined by  three competing interactions: intra polymer, polymer-crowder and crowder-crowder. In addition, the crowder density also plays an important role. The role of attractive crowders on the conformational landscape of a polymer has been explored using coarse grained models for specific crowder densities and specific  crowder-crowder attraction strength~\cite{antypov2008computer,heyda2013rationalizing,huang2021chain}. In an early study~\cite{antypov2008computer}, based on simulations of lattice and off lattice models  of a short polymer chain, it was shown that increasing polymer-crowder attractive interaction leads to a decrease in the radius of gyration, provided the density of the solvent is not too high. These results were obtained for repulsive intra polymer and repulsive  crowder-crowder interactions. Heyda et al~\cite{heyda2013rationalizing} studied the polymer conformations for a range of attractive intra polymer and polymer-crowder interactions. For weak intra polymer interaction, it was shown that increasing polymer-crowder interactions drives a collapse transition due to bridging interactions. In contrast, for strong intra polymer attraction  the configuration was shown to be independent of the polymer-crowder attraction, and no transition from a collapsed state dominated by intra polymer strength to a collapsed state dominated by bridging crowders was reported. For intra polymer interactions, comparable to $\theta $ solvent conditions, it was shown that the polymer initially swells before collapsing due to increased polymer-crowder bridging attraction. These results were obtained for a fixed  crowder-crowder interaction, corresponding to $\theta $ solvent conditions, and also for a fixed crowder density. More recently, Huang et al~\cite{huang2021chain} showed for a single value of strong intra polymer attraction, that the polymer undergoes a collapsed to extended transition followed by an extended to collapsed transition with increasing strength of polymer-crowder attraction. This result is different from earlier obtained results of Heyda et al~\cite{heyda2013rationalizing}.  It is to be noted that the simulations of  Huang et al~\cite{huang2021chain} were done for a stronger  crowder-crowder attraction and higher crowder density as compared to the study by Heyda et al~\cite{heyda2013rationalizing}.

Though the polymer conformations have been explored for some conditions earlier, several issues remain to be clarified. (1) No unified conformational phase diagram in terms of intra polymer and polymer-crowder interaction exists. (2) The role of crowder density and  crowder-crowder attraction on such a phase diagram is not known. (3) Though the concept of bridging crowders has been used to explain the non intuitive collapse of polymer in the presence of attractive crowders, a quantitative definition of the same and using it as an order parameter to delineate phases has not been attempted. (4) In the study by Heyda et al~\cite{heyda2013rationalizing}, there was no transition to a bridging induced collapse phase for a polymer with strong intra polymer interaction, while Huang et al~\cite{huang2021chain} shows such a transition via an intermediate extended phase. It is not  clear whether a direct transition between the two collapsed phases is possible. (5) The observed swelling of the polymer for a certain intra polymer attraction before bridging induced collapse conformation in the study of Heyda et al~\cite{heyda2013rationalizing} was attributed to the $\theta$ solvent conditions. However, a much larger extended conformation was observed by Huang et al~\cite{huang2021chain} for solvent conditions far away from $\theta$ solvent conditions. From these observations, it is not clear whether the observed differences in polymer conformational landscape for strong intra  polymer interactions have their origin in the differences in  crowder-crowder interactions or increased crowder density employed in the earlier studies. 

In this paper, we obtain the conformational phase diagram of a neutral polymer in the intra polymer attraction, crowder polymer attraction phase space.   We identify three phases: $CI$, collapsed phase due to strong intra polymer attraction, with no crowders inside, $E$, the extended phase of the polymer and $CB$, a second collapse phase primarily driven by attractive bridging crowders with significant crowders inside the collapsed phase. The phase diagram is obtained by an analysis of the radius of gyration as well as the number of bridging crowders. The dependence of the phase boundaries on crowder density as well as the  crowder-crowder attraction is clarified. We also show that, in the limit when depletion interactions are dominant,  crowder-crowder attractive interactions can be used to enhance depletion interactions. A discussion is developed to explain the results of earlier work in the context of the phase diagram that is obtained in this paper.

\section{Methods}

We consider a coarse grained bead spring model for a linear, flexible, homopolymer consisting of $N_m$ identical monomers in a box of volume $V=L^3$, in the presence of $N_c$ crowder particles. The neighboring monomers are  connected by harmonic springs. A pair of non bonded particles at distance $r$ interact through the Lennard Jones (LJ) potential
\be
V_{LJ} (r)= 4\epsilon_{ij}\left[ \left(\frac{\sigma_{ij}}{r} \right)^{12}-\left(\frac{\sigma_{ij}}{r} \right)^6 \right],
\ee
where $ij$ can be $mm$, $cc$ and $mc$ depending on whether the pair is monomer-monomer, crowder-crowder or monomer- crowder. We truncate the LJ potential at $r=3 \sigma_{ij}$ for all pairs.  The size of monomers and crowders are taken to be equal ($\sigma_{mm}=\sigma_{cc}=\sigma_{mc}$)

The monomers of the linear polymer are connected via harmonic springs 
\begin{equation}
V_{bond}(r)=\frac{1}{2} k (r-b)^2,
\end{equation}
where $b$ is the equilibrium bond length, $r$ the separation between the bonded particles, and $k$ is the stiffness constant ($k=500 k_BT/\sigma_{mm}^2$). We set  $b=1.12 \sigma_{mm}$. The  bonded neighbors of the polymer are excluded from LJ interactions.  We simulate polymers of length $N_m=50, 100, 150$ though most of the results that are reported are for $N_m=100$, unless otherwise stated.  We simulate systems with different interaction strengths ranging from $\epsilon_{mm}=0.1$--$1.0$ and $\epsilon_{mc}$=$0.1$--$4.0$. For all these systems we simulate two  crowder-crowder interaction strengths: $\epsilon_{mc}=1.0$ and $0.3$. Most of the results are described with $\epsilon_{mc}=1.0$, unless otherwise stated.

The crowder number density, $\rho_c$, is defined as the ratio of number of crowders to the simulation box volume. Most of the simulation results are for crowder number density $\rho_c=0.047$. Higher values of $\rho_c$ are explored to understand the role of crowder density on the conformational phase diagram. The equations of motion are integrated using the MD  LAMMPS software package ~\cite{plimpton1995fast}, and visualization of images and trajectories has been done in visual molecular dynamics (VMD) package \cite{HUMP96}. Various analyses have been performed with in house codes.  All the systems are simulated for $2\times 10^7$ steps using a velocity Verlet algorithm with time step $\delta t=0.001\tau$. The  initial configurations were obtained by setting up a dilute system of neutral polymer  simulated in NPT ensemble for $10^7$ time step to obtain desired density.  Subsequently, simulations are performed under constant volume and temperature condition ($T=1.0$) using Nose Hoover thermostat. Block average over last $9000$ frames is calculated to get the average of each parameter in each simulation. We check that the system reaches equilibrium by doing simulations with both extended  and collapsed initial conditions and verify that the equilibrium configurations are independent of the initial conditions.

\section{Results}

\subsection{\label{sec:phasediagram}Conformational phase diagram with attractive crowders}

In this section, we explore the conformations of a neutral polymer in the $\epsilon_{mm}$--$\epsilon_{mc}$ plane. For all the simulations in this section, crowder number density $\rho_c=0.047$ and  crowder-crowder interaction $\epsilon_{cc}=1.0$. The conformations of the polymer are characterized via its radius of gyration, $R_g$:
\begin{equation}
    R_g^2=\frac{1}{N}\sum_{i=1}^{N}(\textbf{r}_i-\textbf{r}_{\mathrm{cm}})^2,
    \label{eq:rg}
\end{equation}
where $\textbf{r}_i$ is the position vector of the $i$-th monomer and $\textbf{r}_{\mathrm{cm}}$ is the position of center of mass of the polymer.

\begin{figure}
\includegraphics[width=\columnwidth]{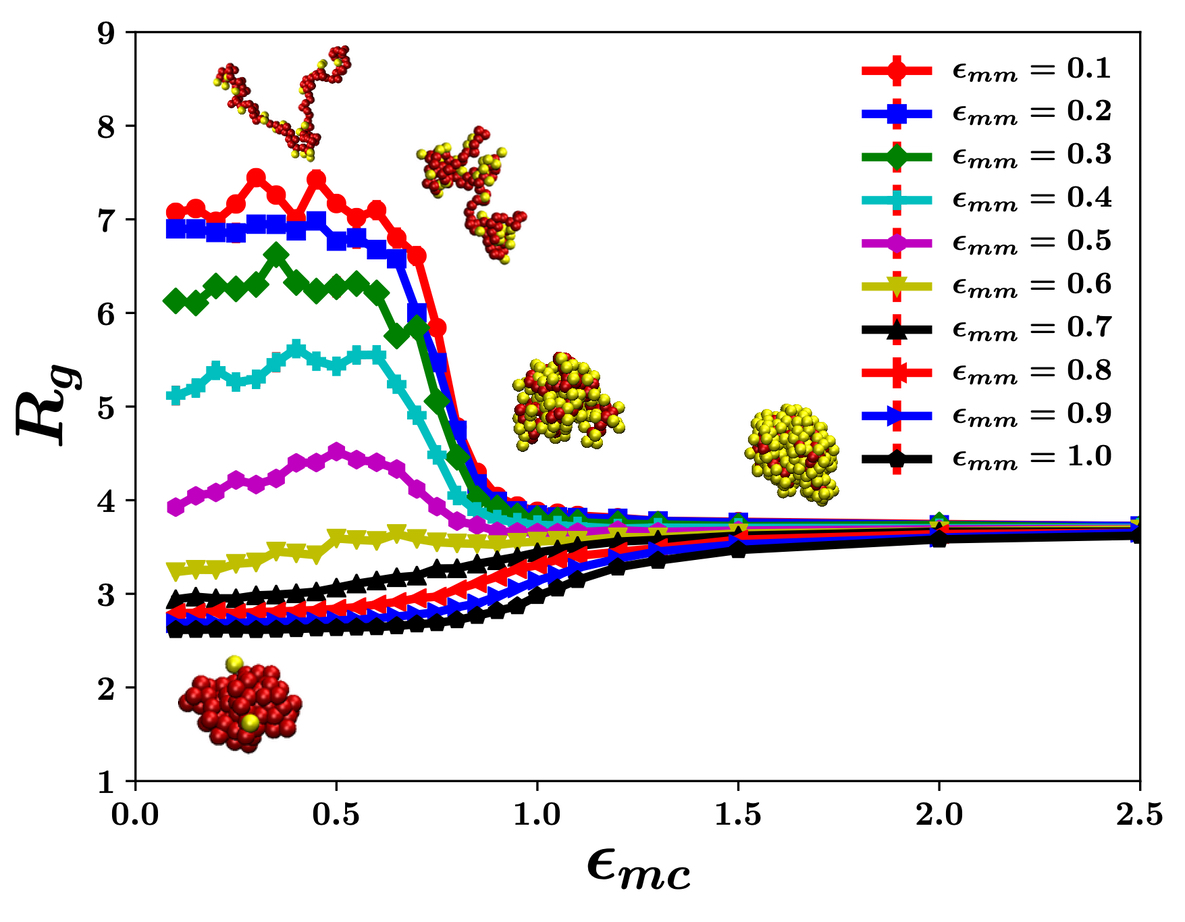}
 \caption{The variation of the mean radius of gyration, $R_g$ with attractive monomer-crowder interaction, $\epsilon_{mc}$,  for different inter monomer interaction, $\epsilon_{mm}$.  In the snapshots along the collapse pathway, monomers are shown in red and crowders that are within a distance of $2\sigma_{mc}$ of at least one monomer  are shown in yellow. All the snapshots are for $\epsilon_{mm}=0.1$ except for the bottom left which is for $\epsilon_{mm}=1.0$. The data are for $\epsilon_{cc}=1.0$ and $\rho_c=0.047$. }
\label{fig:epsilonmcRg}
\end{figure}

Figure~\ref{fig:epsilonmcRg}  shows the variation of $R_g$ with monomer-crowder interaction, $\epsilon_{mc}$, for different values of monomer-monomer interactions, $\epsilon_{mm}$. For small values of $\epsilon_{mc}$, the polymer is extended for weak monomer-monomer interactions ($\epsilon_{mm} \lesssim 0.5$) and collapsed for higher values of $\epsilon_{mm}$. When $\epsilon_{mc}$ is increased, for weak monomer-monomer interactions, the polymer undergoes a transition from an extended phase to a collapsed phase around  $\epsilon_{mc}\approx1.0$ and remains in a collapsed conformation upto the  largest $\epsilon_{mc}$ that we have studied. For intermediate $\epsilon_{mm}$, $R_g$ increases slightly with $\epsilon_{mc}$  before decreasing to a collapsed phase. For  large $\epsilon_{mc}$, regardless of the $\epsilon_{mm}$ values, the polymer is in a collapsed phase. These features, upto $\epsilon_{mc}=1.0$, are as reported in Ref.~\cite{heyda2013rationalizing}. However, in our simulations, we show a second transition from a collapsed phase induced by strong intra polymer attraction to a collapsed phase induced by bridging crowders, when $\epsilon_{mc}$ is increased. We will show that these two collapsed phases differ in their $R_g$ and their structure, aspects that were not explored earlier. 

From these observations, we identify three predominant phases in the conformational phase diagram of neutral polymers with attractive crowders: (1) extended phase, $E$, in which both intra polymer and polymer-crowder  interactions are weak (2) collapsed phase, $CI$, which is characterized by strong intra polymer attraction and (3) collapsed phase, $CB$, which is characterized by bridging interactions due to strong polymer-crowder interactions. We now explore the complete conformational $\epsilon_{mm}-\epsilon_{mc}$ phase diagram and  identify the nature of the three phase lines, $E$--$CB$, $CI$--$CB$ and $E$--$CI$. 

\begin{figure}
\includegraphics[width=\columnwidth]{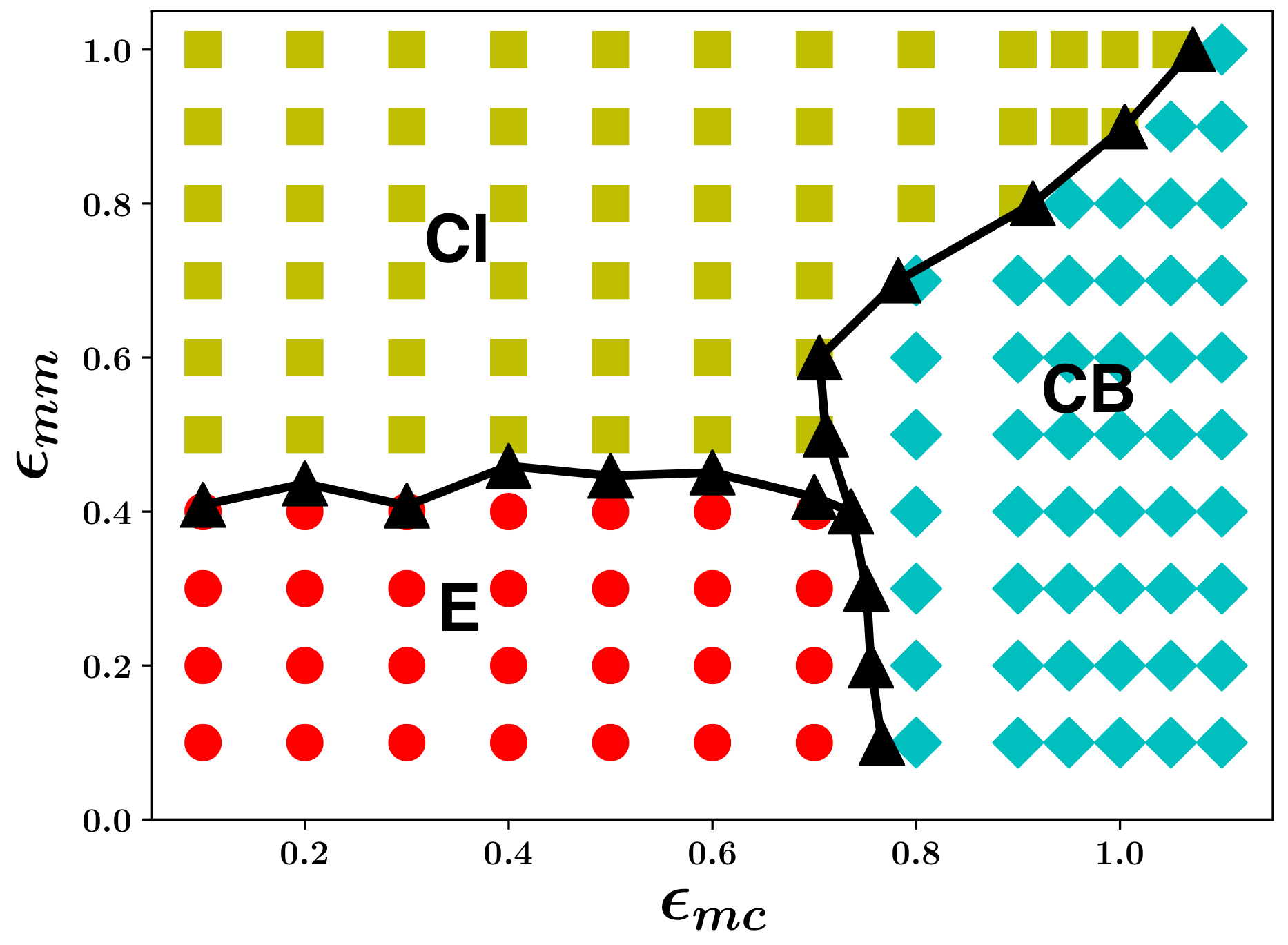}
 \caption{Conformational phase diagram of neutral polymers with attractive crowders in the $\epsilon_{mm}$-$\epsilon_{mc}$ plane. The simulated systems corresponding to extended phase ($E$), strong intra polymer interaction induced collapsed phase ($CI$) and bridging crowder induced collapsed phase ($CB$) are colored in red, yellow and blue respectively. The data are for $\epsilon_{cc}=1.0$ and $\rho_c=0.047$.}
\label{fig:phasediagramEcc1}
\end{figure}

To obtain the phase diagram, we proceed as follows. We identify the transition points between different phases as those values of $\epsilon_{mm}$ and $\epsilon_{mc}$ at which the radius of gyration $R_g$ changes most rapidly, i.e., $dR_g/d\epsilon_{mc}$ or $dR_g/d\epsilon_{mm}$ is maximum. To obtain these gradients, we fit the region near the transition, in Fig.~\ref{fig:epsilonmcRg} to a hyperbolic tangent function. The phase diagram, thus obtained, is shown in Fig.~\ref{fig:phasediagramEcc1}. The three phases extended $E$, collapsed due to intra polymer attraction $CI$, and collapsed due to bridging crowders $CB$, are shown in three different colors. We now examine the polymer conformations and crowder distributions in these three phases to rationalize the phase lines. 

For relatively smaller values of $\epsilon_{mc}$ and moving along the increasing $\epsilon_{mm}$ values, there is a phase transition from extended phase, $E$, to intra polymer attraction dominated collapsed phase, $CI$ at a value of $\epsilon_{mm}^*\approx0.4$, which is consistent with the transition point found in Ref.~\cite{heyda2013rationalizing}. We find that the $\epsilon_{mm}^*$ is utmost weakly dependent on $\epsilon_{mc}$, which may be understood as follows. For the low value of crowder density simulated, the typical monomer-crowder separation ($\rho_c^{-1/3}\approx 2.77$) being larger than the LJ cutoff, the $E$--$CI$ transition is largely independent of $\epsilon_{mc}$.  

For $\epsilon_{mm}=0.1$, the $E$--$CB$ transition occurs at $\epsilon_{mc}^*\approx 0.77$. For higher values of $\epsilon_{mm}$, the phase line slopes slightly to the left (the slope being larger when $\epsilon_{cc}$ is lowered, as in Sec.~\ref{sec:epsiloncc}) showing that bridging interactions start dominating at a smaller value of $\epsilon_{mc}$.  The decrease of $\epsilon_{mc}^*$ with increasing $\epsilon_{mm}$ can be rationalized as follows. Since the intra polymer interactions are attractive, an increase in $\epsilon_{mm}$ enhances the propensity to collapse. An additional inclusion of bridging crowder interaction ($\epsilon_{mc}$) only further stabilizes the $CB$ phase and hence lowers the critical value of $\epsilon_{mc}^*$ needed for the transition.  

In contrast, beyond $\epsilon_{mm}\approx0.6$, when the polymer is in $CI$ phase for small $\epsilon_{mc}$,  $\epsilon_{mc}^*$ increases with  $\epsilon_{mm}$ for the $CI$--$CB$ transition. This can be rationalized as follows. In the $CI$ phase, every non-bonded monomer-monomer interaction lowers energy by $\epsilon_{mm}$. The $CI$--$CB$ transition requires the replacement of a strong monomer-monomer interaction by at least an equivalent monomer-crowder interaction, resulting in higher $\epsilon_{mc}$ values for larger $\epsilon_{mm}$. 
\begin{figure}
\includegraphics[width=\columnwidth]{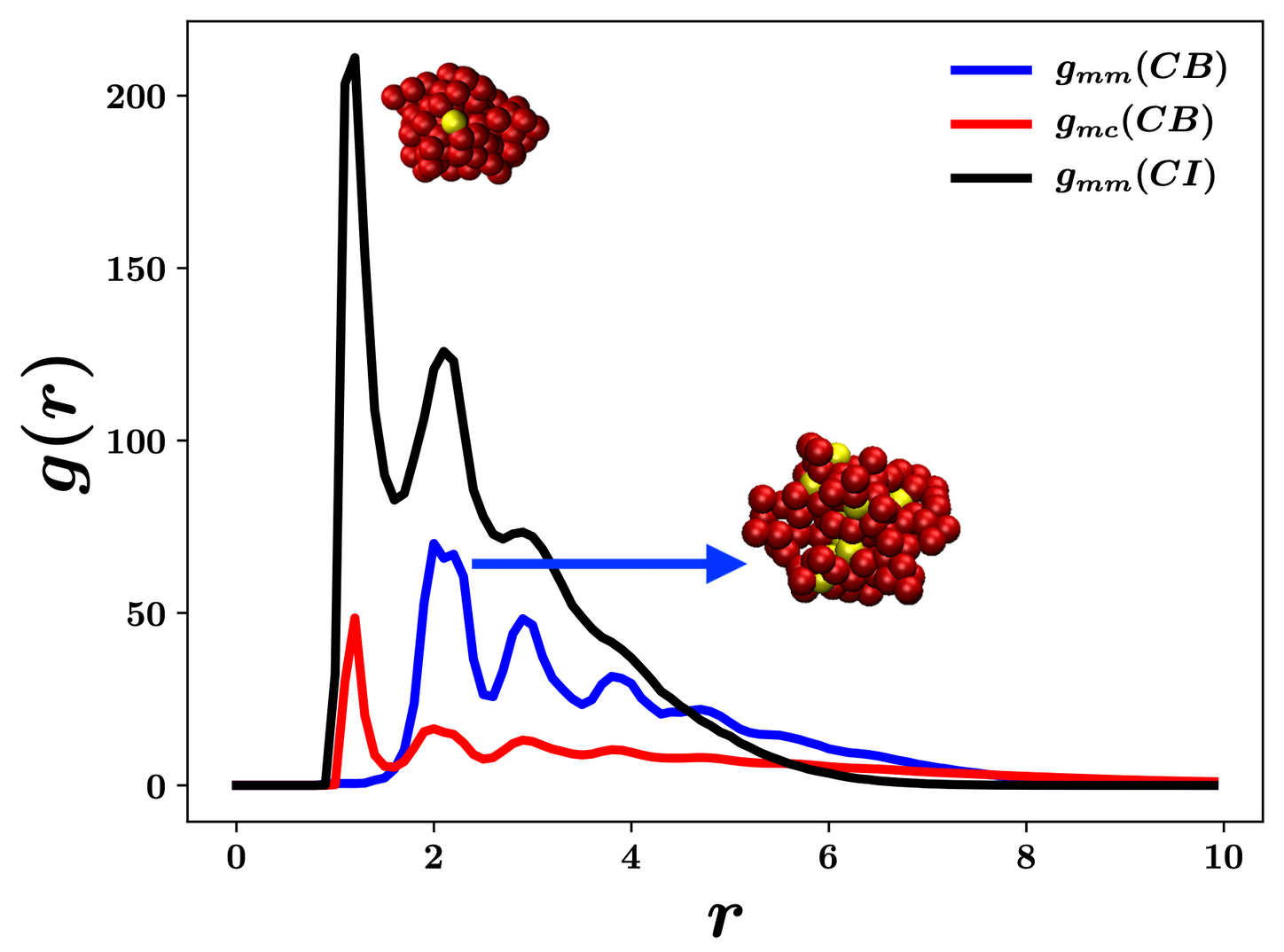}
 \caption{The pair correlation function, $g(r)$, for both monomer-monomer and monomer-crowder in the $CB$ phase ($\epsilon_{mm}=0.1$, $\epsilon_{mc}=4.0$) and $CI$ phase ($\epsilon_{mm}=1.0$, $\epsilon_{mc}=0.1$). The data are for $\epsilon_{cc}=1.0$ and $\rho_c=0.047$. The contributions from the bonded monomers is excluded. Snapshots of the polymer in $CI$ and $CB$ phases are shown in red with bridging crowders (shown in yellow).} 
\label{fig:gr-epsilonmm}
\end{figure}

We now quantitatively differentiate between the two collapsed phases $CI$ and $CB$. To do so, we compare the monomer-monomer pair correlation function, $g(r)$, for these two phases in Fig.~\ref{fig:gr-epsilonmm}. In the calculation of $g(r)$, the bonded pairs are excluded. In the $CI$ phase, there is a peak close to $r\approx1.1$, corresponding to the large number of non-bonded monomer-monomer pairs leading to the collapsed phase. In sharp contrast, this peak is completely absent in the $CB$ phase, showing that intra polymer interactions are less significant. In the $CB$ phase, the dominant interactions are monomer-crowder as can be seen from the peak at $r\approx 1.1$ in the monomer-crowder $g_{mc}(r)$ in  Fig.~\ref{fig:gr-epsilonmm}. Further, the larger extent of $g_{mm}(r)$ for the $CB$ phase, as compared to the $CI$ phase, is consistent with a larger $R_g$. In addition, due to the dominant intra polymer interactions in the $CI$ phase, we expect $g_{mm}(r)$ to depend strongly on $\epsilon_{mm}$. However, when the polymer-crowder interactions are dominant, as in the $CB$ phase, we expect $g_{mm}(r)$ to be largely independent of $\epsilon_{mm}$. This aspect is clearly seen in Fig.~S1 (see Supp Info), where  $g_{mm}(r)$ for different $\epsilon_{mm}$ are shown. For $\epsilon_{mc}=0.1$, corresponding to $CI$ phase, $g_{mm}$ has a strong dependence on the $\epsilon_{mm}$ and the emergence of a structure between monomers can be seen as $\epsilon_{mm}$ is increased [see Fig.~S1(a)]. However, for strong polymer-crowder attraction, the structure of the collapsed phase $CB$ is identical within numerical error [Fig.~S1(b)] and has a similar structure corresponding to a collapsed phase. In addition, the overall size of the collapsed state  $CB$ approaches a limiting value (independent of $\epsilon_{mm}$ and function of only polymer length). We conclude that the $CB$ phase is distinctly different from the  $CI$ phase and largely independent of intra polymer interactions.

\begin{figure}
\includegraphics[width=\columnwidth]{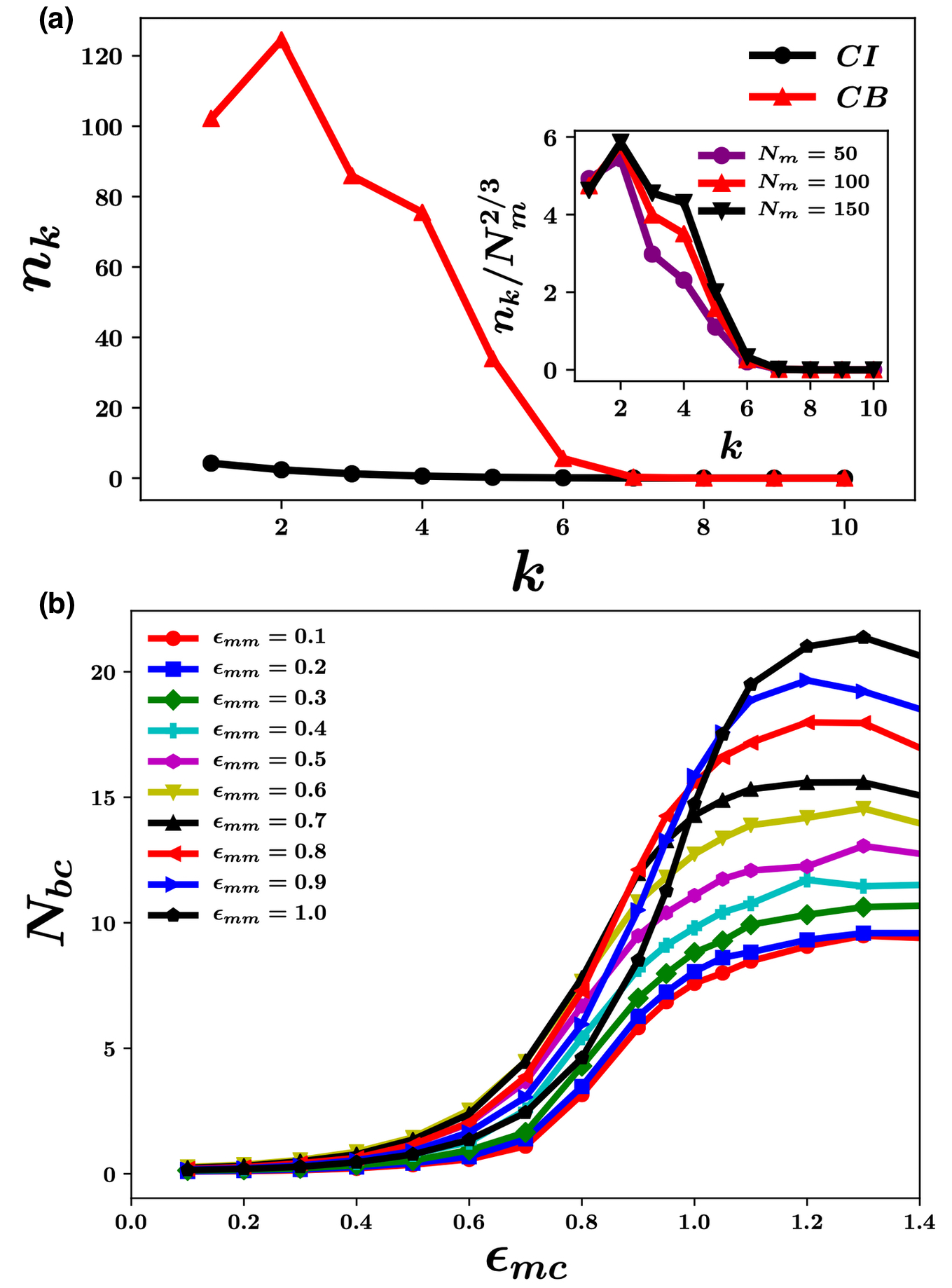}
 \caption{(a) The variation of $n_k$, the number of  crowders with $k$ monomers within a radius of $1.5\sigma$, with $k$ in the $CI$ ($\epsilon_{mm}=0.9$, $\epsilon_{mc}=0.2$)and $CB$ ($\epsilon_{mm} =1.0$, $\epsilon_{mc}=4.0$) phases. Inset: The data for small $k$ and difference polymer lengths $N_m$ collapse onto a single curve when $n_k$ is scaled by $N_m^{2/3}$. (b) The number of bridging crowders, $N_{bc}$ [see Eq.~(\ref{eq:nbc})],  representing an order parameter across $E-CB$ and $CI-CB$ phase lines, for different values of $\epsilon_{mm}$.  }
 \label{fig:histNk}
\end{figure}

We now look at bridging crowders to further differentiate between the $CB$ and $CI$ collapsed phases. Although the concept of attractive crowders acting as bridges or glue between distant monomers has been invoked in earlier works~\cite{heyda2013rationalizing,rodriguez2015mechanism,huang2021chain,ryu2021bridging,brackley2020polymer,brackley2013nonspecific}, a quantitative definition of the same is lacking. We define a crowder as a bridging crowder if it interacts with at least $k$ monomers.  It is apriori not clear what the optimal value of $k$ should be. If $k$ is too small, the crowders that are adsorbed on the surface of the collapsed polymer will be incorrectly counted as bridging crowders. On the other hand, if $k$ is too large, very few crowders will satisfy the bridging criterion. We determine the optimal value of $k$ as follows. Let $n_k$ denote the number of crowders that have exactly $k$ monomers within a sphere of radius $1.5 \sigma$ centered about the crowder. We choose $1.5 \sigma$ as it is larger than the LJ minimum of $2^{1/6} \sigma$ and smaller than the position of the second peak seen in the monomer-crowder pair correlation function (see Fig.~\ref{fig:gr-epsilonmm}). The variation of $n_k$ with $k$ is shown in Fig.\ref{fig:histNk}(a) for parameter values corresponding to both $CI$ and $CB$ phases. We find that for the $CI$ phase, $n_k$ is nearly zero for $k>2$. On the other hand, for the CB phase, we find that $n_k$ is comparable to the length of the polymer for small values of $k$ and decreases to zero for $k>10$. To distinguish surface adsorbed crowders from bridging crowders, we calculate $n_k$ for three different values of polymer length $N_m$. We find that the data for small $k$ and different $N_m$ collapse onto one curve when $n_k$ is scaled by  $N_m^{2/3}$ [see inset of Fig.\ref{fig:histNk}(a)]. Since the surface area of a collapsed polymer scales as  $N_m^{2/3}$, crowders with small values of $k$ correspond to surface adsorbed crowders. This puts a lower bound ($k>3$) for the value of $k$ for defining a bridging crowder.

We select $k=6$ as the definition of a bridging crowder (also see argument below), i.e., a crowder which is within $1.5\sigma$ distance from at least 6 monomers. The number of such bridging crowders, $N_{bc}$, is then calculated from the data in Fig.~\ref{fig:histNk}(a) as 
\be
N_{bc}=\sum_{k=6}^\infty n_k. 
\label{eq:nbc}
\ee
The variation of $N_{bc}$ with $\epsilon_{mc}$ is shown in Fig.~\ref{fig:histNk}(b) for different values of  $\epsilon_{mm}$. Around the value of  $\epsilon_{mc}^*$, as calculated in Fig.~\ref{fig:phasediagramEcc1}, $N_{bc}$ increases sharply from zero to non-zero value and thus can be used as an order parameter to distinguish phases with and without bridging crowders. We calculate $\epsilon_{mc}^*$ from the $N_{bc}$ data following the same procedure as was done with $R_g$ data, and the phase diagram, thus obtained, is shown in Fig.~S2 (see Supp Info).  We find the phase diagram obtained using $R_g$ and $N_{bc}$ are qualitatively the same, thus under scoring the robustness of the identification of the phases and the phase diagram in Fig.~\ref{fig:phasediagramEcc1}. We have constructed the phase diagram for $k=4$--$7$ and we find the phase diagram with $k=6$ to be the closest to the one obtained from the analysis of $R_g$.

\subsection{Effect of crowder-crowder interaction $\epsilon_{cc}$ on the conformational phase diagram \label{sec:epsiloncc}}

\begin{figure}
\includegraphics[width=\columnwidth]{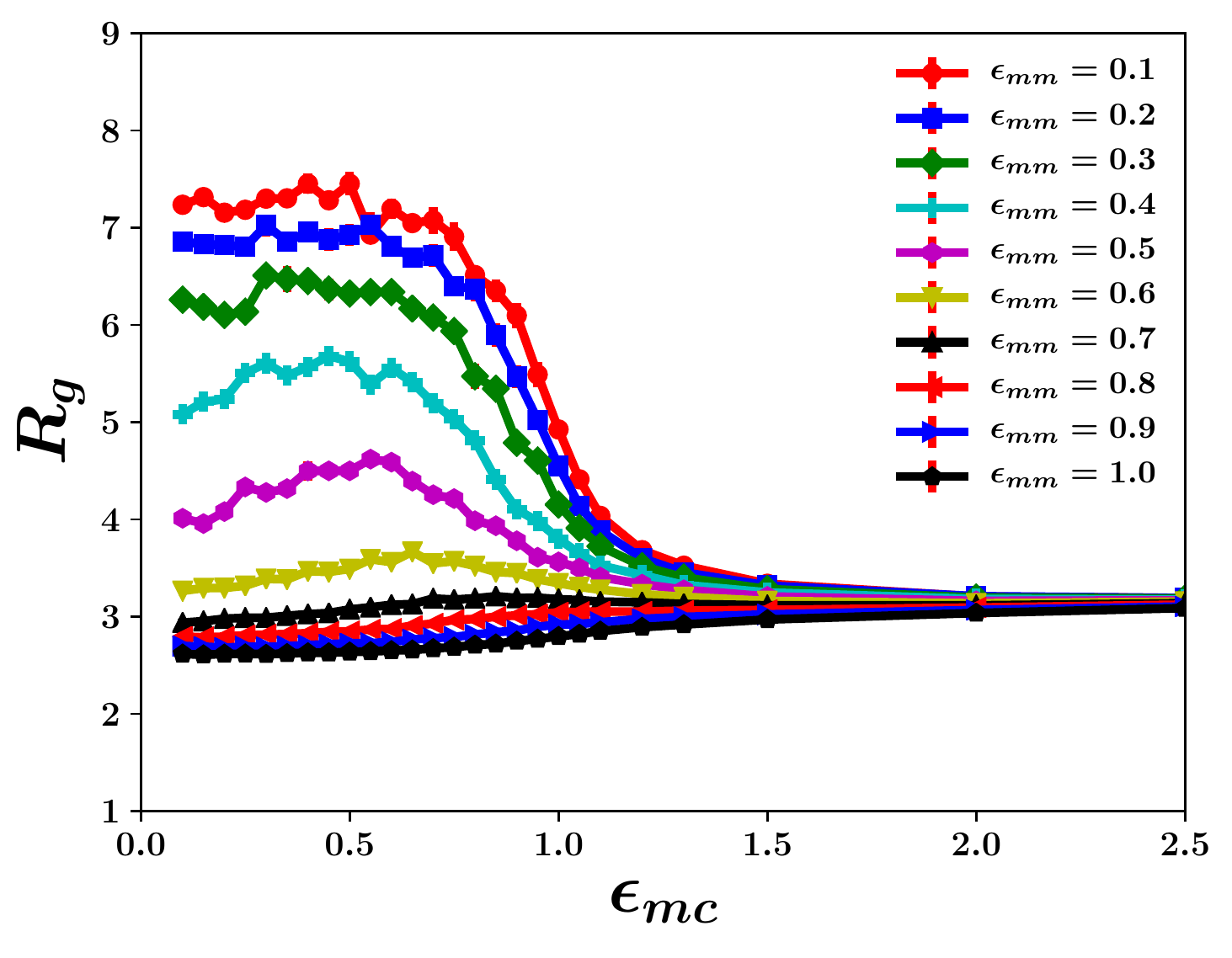}
 \caption{The variation of the mean radius of gyration, $R_g$ with attractive monomer-crowder interaction, $\epsilon_{mc}$,  for different inter-monomer interaction, $\epsilon_{mm}$ for reduced crowder-crowder interaction, $\epsilon_{cc}=0.3$.  In the snapshots along the collapse pathway, monomers are shown in red and crowders that are within a distance of $2\sigma_{mc}$ of at least one monomer  are shown in yellow. All the snapshots are for $\epsilon_{mm}=0.1$ except for the bottom left which is for $\epsilon_{mm}=1.0$. The data are for $\rho_c=0.047$. }
\label{fig:epsilonmcRg_2}
\end{figure}
In this section we ask how the conformational phase diagram obtained in Fig.~\ref{fig:phasediagramEcc1} is affected by the crowder-crowder attractive interactions, $\epsilon_{cc}$.  Towards this, we obtain the conformational phase diagram for a lower value of $\epsilon_{cc}=0.3$, as chosen in Heyda et al~\cite{heyda2013rationalizing}. The crowder density is kept fixed at $\rho_c=0.047$.
\begin{figure}
\includegraphics[width=0.85\columnwidth]{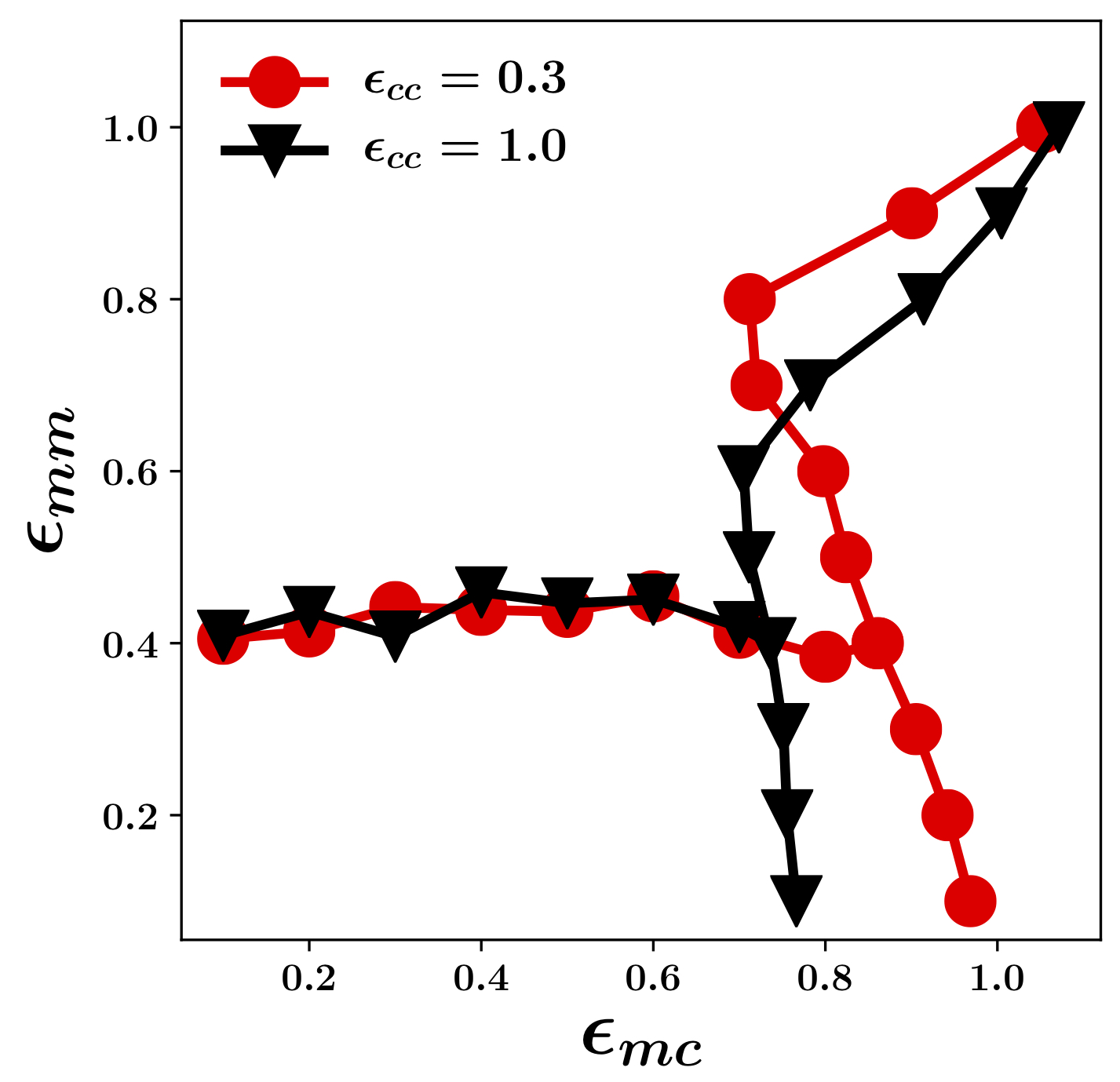}
 \caption{The conformational phase diagram of the polymer in the $\epsilon_{mm}$-$\epsilon_{mc}$ plane for $\epsilon_{cc}=0.3, 1.0$. The data are for $\rho_c=0.047$.}
 \label{fig:phasediagram2}
\end{figure}

Figure~\ref{fig:epsilonmcRg_2} shows the variation of $R_g$ with monomer-crowder interaction, $\epsilon_{mc}$, for different values of monomer-monomer interactions, $\epsilon_{mm}$ and for $\epsilon_{cc}=0.3$. Though the qualitative features of $R_g$ are similar for both $\epsilon_{cc}=0.3$ and $\epsilon_{cc}=1.0$ (see Fig.~\ref{fig:epsilonmcRg}), there is a dependence of the transition points as well as the collapsed polymer size on $\epsilon_{cc}$. We follow the same procedure, as described in Sec.~\ref{sec:phasediagram} to identify the phase diagram. The phase diagrams for the two values of  $\epsilon_{cc}$ are compared in Fig.~\ref{fig:phasediagram2}. The $E$--$C1$ phase line is not affected by $\epsilon_{cc}$.  This is because, for this range of $\epsilon_{mc}$ values and the crowder density $\rho_c$, the polymer-crowder interaction is not significant and hence crowder-crowder interaction plays no role. The data also suggest that, when $\epsilon_{cc}$ is decreased, then  the $E$--$CB$ transition occurs at a larger value of $\epsilon_{mc}$. This dependence is also clearly evident in Fig.~S3(a) (see SuppInfo), where $R_g$ for two values of $\epsilon_{cc}$ are compared for weak ($\epsilon_{mm}=0.1$)  intra polymer attraction. In contrast, the phase diagram also shows that the $CI$--$CB$ transition occurs at a smaller value of  $\epsilon_{mc}$ for smaller values of $\epsilon_{cc}$. This is also seen in  Fig.~S3(b) (see SuppInfo), where $R_g$ for two values of $\epsilon_{cc}$ are compared for strong ($\epsilon_{mm}=1.0$). Hence, the effect of the crowder-crowder interaction is opposite for small and large $\epsilon_{mm}$ values and is reflected in the crossover of the phase lines with increasing $\epsilon_{mm}$. 

The effect of $\epsilon_{cc}$ on the $E$--$CB$ and $CI$--$CB$ phase boundaries can be understood as follows.  Consider intermediate values of $\epsilon_{mc}$. Keeping other parameters fixed, if $\epsilon_{cc}$ is increased then the effective good solvent nature is diminished because increasing $\epsilon_{cc}$ enhances crowder-crowder interaction and effectively reduces monomer-crowder interactions.  Applying this observation to the $E$ phase, we conclude that the $E$ phase, in which the polymer is in an effectively good solvent condition, becomes less stable when $\epsilon_{cc}$ is increased at intermediate $\epsilon_{mc}$. This implies that we need a smaller value of $\epsilon_{mc}$  to induce the $CB$ phase, resulting in the leftward shift of the $E$--$CB$ phase boundary. 
To understand the rightward shift in the $CI$--$CB$ with increasing $\epsilon_{cc}$, we proceed as follows. In the $CI$ phase, the polymer is in an effective poor solvent condition characterized by high values of $\epsilon_{mm}$. When $\epsilon_{cc}$ is increased, based on the above observation, the effective poor solvent condition is enhanced. Thus, the system has an effectively higher $\epsilon_{mm}$, and since $\epsilon_{mc}^*$ increases with $\epsilon_{mm}$, we obtain that a larger value of $\epsilon_{mc}^*$ is needed for the  $CI$--$CB$ transition. These arguments also explains the cross over of the phase lines  seen in Fig.~\ref{fig:phasediagram2}.

\subsection{Effect of crowder density on neutral polymer conformational phase diagram}

The density of the crowders can have strong effects on the polymer conformation. In the well known example of depletion-induced interactions, following the pioneering work of Asakura and Oosawa~\cite{asakura1954interaction,asakura1958interaction,miyazaki2022asakura}, it is known that purely repulsive interactions between the polymer and crowders can drive the collapse of the polymer. The depletion interactions are enhanced by increasing crowder density as well as decreasing crowder size. It is, thus, reasonable to expect that density will play an important role in bridging interactions resulting from attractive crowder-polymer and crowder-crowder interactions, as studied here.

\begin{figure}
\includegraphics[width=0.85\columnwidth]{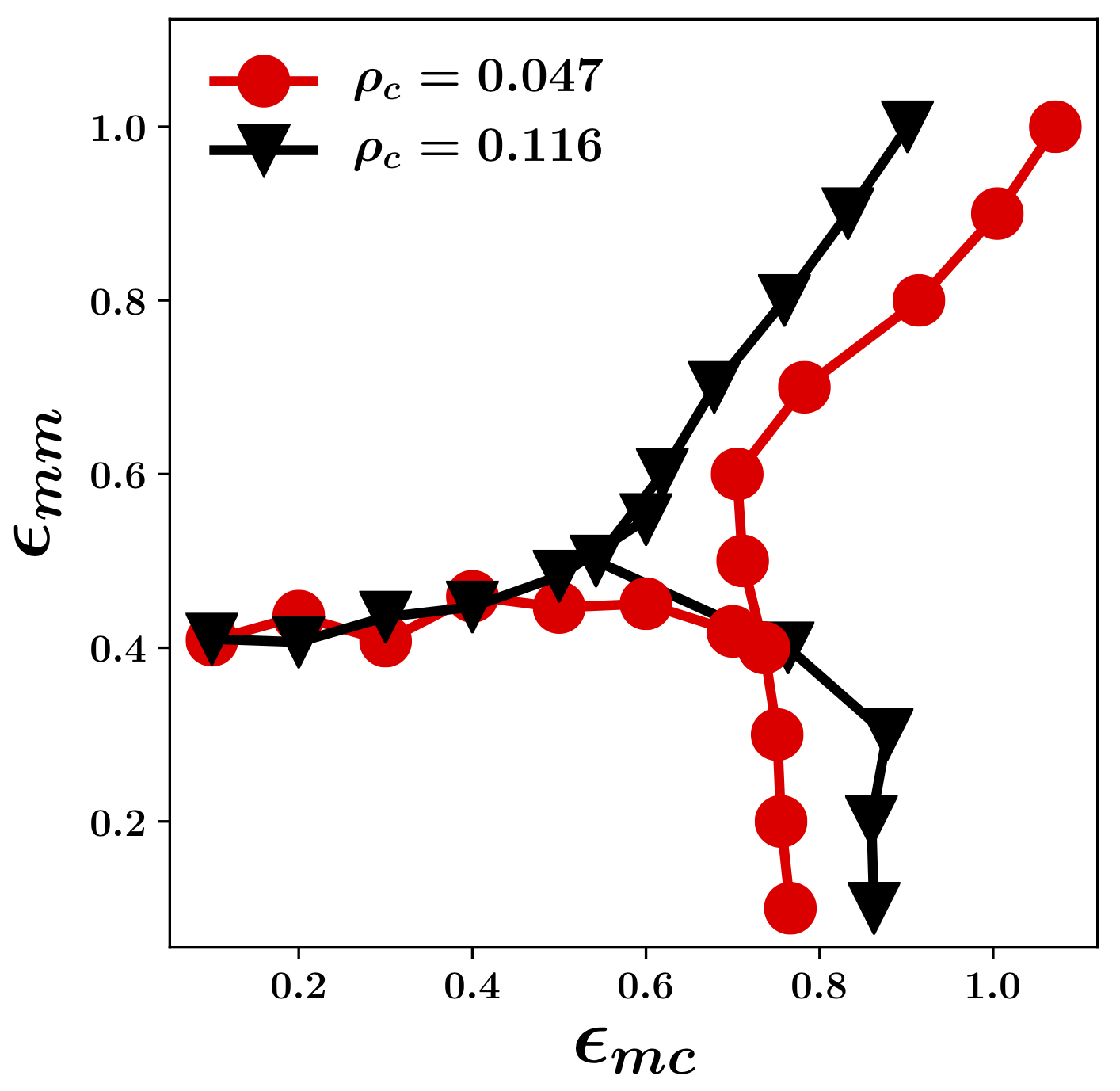}
 \caption{The conformational phase diagram of the polymer in the $\epsilon_{mm}$-$\epsilon_{mc}$ plane for $\rho_c=0.047$ and $0.116$. The data are for $\epsilon_{cc}=1.0$.}
\label{fig:phasediagramphi}
\end{figure}

To probe the role of crowder density $\rho_c$, we do simulations for a different crowder density $\rho_c=0.116$. This value of crowder density coincides with that in Heyda et al~\cite{heyda2013rationalizing}. The phase diagram obtained for this density, following the same procedure as  described in Sec.~\ref{sec:phasediagram}, is compared to that for $\rho_c=0.047$ in Fig.~\ref{fig:phasediagramphi}.

Changing  $\rho_c$ has a strong effect on the phase boundaries. When $\rho_c$ is increased,  the $E$--$CI$ phase boundary slopes upwards more with increasing $\epsilon_{mc}$. This should be contrasted with the insignificant role played by $\epsilon_{cc}$ on the $E$--$CI$ phase boundary [see Fig.~\ref{fig:phasediagram2}]. The $E$--$CB$ phase boundary shifts to the right with increasing $\rho_c$ while the $CI$--$CB$ shifts to the left with increasing $\rho_c$. These opposing effects at small and large $\epsilon_{mm}$ result in a crossover of the phase lines. Further evidence of these shifts of the $E$--$CI$ and $E$--$CB$ phase boundaries can be seen in Fig.~S4 (see Supp Info), where the variation of $R_g$ with $\epsilon_{mc}$ is compared for different $\rho_c$ for weak intra polymer interaction ($\epsilon_{mm}=0.1$) and strong intra polymer interaction ($\epsilon_{mm}=1.0$). 
When $\epsilon_{mm}=0.1$, $R_g$ decreases sharply at a lower value of $\epsilon_{mc}$ for $\rho_c=0.047$, consistent with the phase diagram in Fig.~\ref{fig:phasediagramphi}. However, for $\epsilon_{mm}=1.0$, $R_g$ increases sharply at a smaller value of $\epsilon_{mc}$ for $\rho_c=0.116$, again consistent with the phase diagram in Fig.~\ref{fig:phasediagramphi}.

The effect of $\rho_c$ on the phase boundaries can be rationalized, based on one key observation. At intermediate values of $\epsilon_{mc}$, increasing $\rho_c$ enhances the effective good solvent nature, due to increase in number of crowders in the vicinity of the polymer. Applying this observation to the $E$ phase, in which the polymer is in an effectively good solvent condition, we conclude that the polymer prefers more to be in $E$ phase when $\rho_c$ is increased at intermediate $\epsilon_{mc}$. This implies that a larger value of $\epsilon_{mm}$ is needed at intermediate $\epsilon_{mc}$ to induce the $CI$ phase, thus explaining the shift in the $E$--$CI$ phase boundary. The same preference of the polymer to be in $E$ phase also implies that a larger value of $\epsilon_{mc}$ is required to induce the $CB$ phase, explaining the shift in the $E$--$CB$ phase boundary with changing $\rho_c$. To understand the leftward shift in the $CI$--$CB$ with increasing $\rho_c$, we proceed as follows. In the $CI$ phase, the polymer is in an effective poor solvent condition characterized by high values of $\epsilon_{mm}$. When $\rho_c$ is increased, based on the above observation, the effective poor solvent condition is weakened. Thus, the system has an effectively lower $\epsilon_{mm}$, and since $\epsilon_{mc}^*$ decreases with $\epsilon_{mm}$, we obtain that a smaller value of $\epsilon_{mc}^*$ is needed for the eventual $CI$--$CB$ transition. These arguments also explains the cross over of the phase lines  seen in Fig.~\ref{fig:phasediagramphi}. 

In addition, when crowder density $\rho_c$ is increased, there is a significant difference in the $R_g$ values in the $CB$ phase (see Fig.~S4 in Supp Info), likely due to increased number of bridging crowders present inside the collapsed phase. Only when $\epsilon_{mc}$ is significantly increased, the $R_g$ at $\rho_c=0.116$ approaches the limiting value of $R_g$ at $\rho_c=0.047$.  Though the crowder density $\rho_c=0.116$ is same as that of Ref.~\cite{heyda2013rationalizing}, the sharp increase of $R_g$ value for $CI$ phase seen here was not observed in the earlier study. This is likely due to smaller value of $\epsilon_{cc}=0.3$ used in that study. Huang et al~\cite{huang2021chain} observed a $CI$--$E$--$CB$ phase, which we do not see in our study. Though the $\epsilon_{cc}=1.0$ used in this study is same, a much higher density of $\rho_c=0.64$ was used in that study. From the results seen in Fig.~S4 (see Supp Info), we can extrapolate that a further increase in  $\rho_c$ would likely show the same result as that of Huang et al~\cite{huang2021chain}. These results demonstrate conclusively that the apparently inconsistent results in the polymer conformations in the previous simulations~\cite{heyda2013rationalizing,huang2021chain} can be explained in terms of crowder density and crowder-crowder interactions. The results also show that the conformational landscape of neutral polymer is significantly more nuanced in the presence of crowders (especially attractive crowders) than a simple binary picture of good/bad solvent conditions leading to extended/collapsed conformations.

\begin{figure}
\includegraphics[width=\columnwidth]{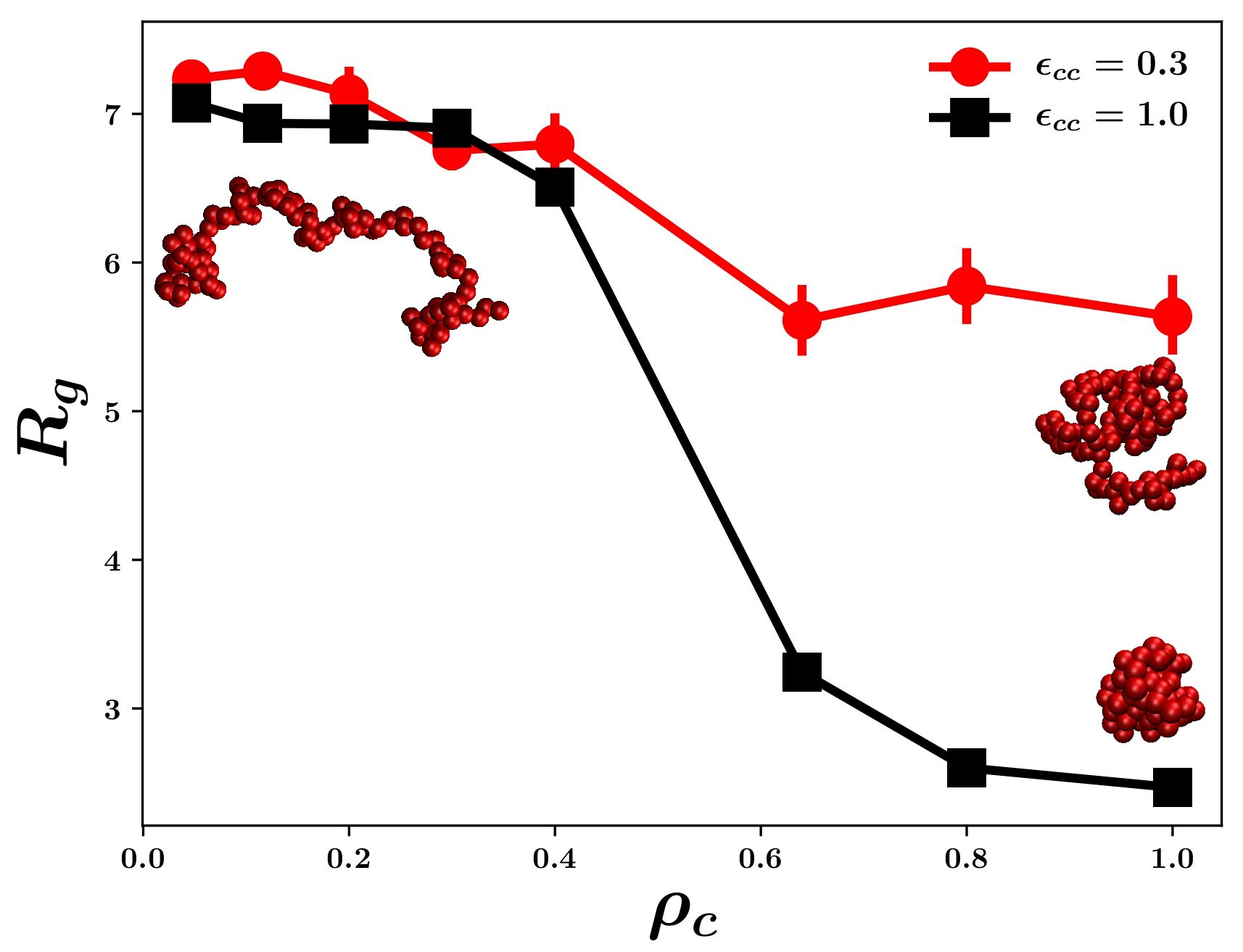}
 \caption{Polymer conformation dependence, in terms of radius of gyration $R_g$, on crowder density, $\rho_c$, and crowder-crowder interaction, $\epsilon_{cc}$, for weak intra polymer interaction $\epsilon_{mm}=0.1$ and weak polymer-crowder interaction $\epsilon_{mc}=0.1$. Snapshots of typical configurations of the polymer are also shown for different crowder densities.}
\label{fig:phicompareweak}
\end{figure}

We now further explore the interplay between crowder density, $\rho_c$, and crowder-crowder interactions, $\epsilon_{cc}$. Results in Sec.~\ref{sec:epsiloncc} show that when $\epsilon_{cc}$ is increased, the effective poor solvent condition is enhanced. In contrast, increasing $\rho_c$ (in the range where inter particle separation is much larger that LJ minimum) results in an  enhanced effective good solvent conditions. We now probe 
the combined effect on the polymer conformations  at larger values of $\rho_c$ than considered in Fig.~\ref{fig:phasediagramphi}. We use the extended structure (seen for  $\epsilon_{mm}=0.1$ and $\epsilon_{mc}=0.1$) as the starting configuration for these simulations and the results are shown in Fig.~\ref{fig:phicompareweak}. For $\epsilon_{cc}=1.0$, we see a sharp transition from extended to a collapsed phase as $\rho_c$ is increased beyond $\rho_c=0.6$. In contrast, for smaller value of $\epsilon_{cc}=0.3$ the polymer remains in an extended state with slight reduction of $R_g$ as $\rho_c$ is increased beyond $\rho_c=0.6$. These results should be compared with the compaction achieved by purely repulsive crowders in depletion induced collapse transition ~\cite{asakura1954interaction,asakura1958interaction,miyazaki2022asakura,kang2015effects}, unlike the attractive crowders used in this study. Further, it has been shown that the size of the crowders can enhance such depletion induced collapse. To achieve a compaction of $\approx 66\%$, a crowder size of $1/10$th the monomer size was required~\cite{kang2015effects}. Here we show that the same compaction can be achieved with the crowder size same as the monomer size but with enhanced crowder-crowder attraction. These results also suggest that in addition to the two collapsed phases discussed in this study, $CI$ (strong intra polymer attraction) and $CB$ (strong polymer-crowder attraction), an additional density induced collapsed state for enhanced crowder-crowder interaction can be seen: density induced collapse phase, $CD$.

\section{Summary and Discussion}

In this paper, we determine the phase diagram in the intra polymer attraction ($\epsilon_{mm}$) and polymer-crowder attraction ($\epsilon_{mc}$) phase space, and show explicitly the existence of three distinct phases: collapsed phase $CI$, primarily driven by large $\epsilon_{mm}$, extended phase $E$ for small $\epsilon_{mm}$ and $ \epsilon_{mc} $, and a second collapsed phase $CB$,  driven primarily by bridging induced attraction for large $\epsilon_{mc}$. The phase boundaries delineating the three different phases are obtained by identifying the inflection point in the data for the radius of gyration, $R_g$.  By exploring a wide range of intra polymer and polymer-crowder interactions, we conclusively show the existence of a  $CI$--$CB$ phase transition, which was hitherto unexplored.  This is in addition to the $E$--$CB$ and $E$--$CI$ transitions, which were known earlier. The measured $R_g$ values, along with the structurally informative pair correlation functions suggest that the bridging induced collapsed phase $CB$ is characterized purely by polymer-crowder interaction and is insensitive to intra polymer interaction strength. We also quantify the definition of a bridging crowder by computing the number of monomers within a cutoff radius of a crowder. We show that the number of bridging crowders can be used as an order parameter to differentiate the bridging induced phase $CB$ from the $E$ and the $CI$ phases. The conformational phase diagram constructed using this order parameter is shown to be similar to the phase diagram obtained via analysis of $R_g$.  The effect of crowder density $\rho_c$ and crowder-crowder interaction, $\epsilon_{cc}$, on the phase boundaries between different phases is also explored. In particular the $E$--$CB$ transition occurs at a larger value of $\epsilon_{mc}$ with increasing $\rho_c$ and at a smaller value of $\epsilon_{mc}$  with increasing $\epsilon_{cc}$. On the other hand, the effect is the opposite for the $CI$--$CB$ transition. The $CI$--$CB$ transition occurs at a smaller value of $\epsilon_{mc}$ with increasing $\rho_c$ and at a larger value of $\epsilon_{mc}$  with increasing $\epsilon_{cc}$. We provide rationalizations for these effects in the paper.

Our result for the phase diagram encompasses known earlier results for bridging interactions by attractive crowders~\cite{heyda2013rationalizing,huang2021chain}, and accounts for the difference in polymer conformations seen in these works.  When $\epsilon_{mm}$ and $\epsilon_{mc}$ are comparable, the polymer adopts a slightly swollen phase (not extended) before undergoing a collapse transition at higher $\epsilon_{mc}$ in Heyda et al~\cite{heyda2013rationalizing}, whereas the polymer adopts a fully extended state before the bridging-induced collapse with increasing $\epsilon_{mc}$ in Huang et al~\cite{huang2021chain}. It is to be noted that the crowder density $\rho_c$ and crowder-crowder interaction $\epsilon_{cc}$ are both different in the two studies. While Heyda at al~\cite{heyda2013rationalizing} use a lower density ($\rho_c=0.116$) and smaller $\epsilon_{cc}=0.3$, Huang  et al~\cite{huang2021chain} does simulations at significantly higher crowder density ($\rho_c=0.64$) and larger $\epsilon_{cc} =1.0$. In our phase diagram, as $\rho_c$ is increased, the $E$--$CI$ phase boundary starts developing an upward slope, i.e., when $\epsilon_{mc}$ is increased the $E$--$CI$ transition occurs at a larger value of $\epsilon_{mm}$. This effect is more pronounced at higher crowder densities like $\rho_c=0.64$ in Huang  et al~\cite{huang2021chain}, which can also be extrapolated from our data. In such a situation when $\epsilon_{mc}$ is increased, there is a range of $\epsilon_{mm}$ for which an extended phase can appear between the $CI$ and $CB$ phases, leading to $CI$--$E$--$CB$ sequence of transitions as seen in Huang  et al~\cite{huang2021chain}.

Our results strongly indicate that the interplay between  crowder-crowder interactions and crowder density can have non-intuitive consequences to the neutral polymer conformational phase diagram. The usual understanding of the role of crowders on the conformations of a neutral polymer has been through the lens of entropic volume exclusion~\cite{minton1981excluded,zhou2008macromolecular,gnutt2015excluded}. It is to be noted that in the much celebrated Asakura-Oosawa model~\cite{asakura1958interaction},  which is the basis for such volume exclusion dominated macromolecular collapsed conformations, the core assumption is that the crowders and macromolecule interact via only purely steric or repulsive interactions. In our study, we explore the role of crowder density on the polymer conformations with two important differences with depletion model: crowders have attractive interactions with polymer and with other crowders. Our results show that neutral polymers with weak intra polymer attraction can also undergo a strong extended to collapse transition with attractive crowder density and this can be modulated by  crowder-crowder attraction. This result can have significant relevance for studies which suggest that depletion induced stability picture ignores the enthalpic considerations via direct interactions between the crowders and the macromolecules~\cite{senske2014protein,politi2010enthalpically,sukenik2013balance,sapir2014origin,sapir2015depletion,miklos2011protein,guin2019weak,banks2018intrinsically}. 
There have also been studies which seek to explain genome organization and biological condensates via simple polymer models~\cite{brackley2013nonspecific,ancona2022simulating,brackley2020polymer,ryu2021bridging,barbieri2012complexity}, both lattice and continuum. Many of these studies do not take into account the  crowder-crowder attractive interactions which are very relevant in the crowded cellular environment and the results obtained here could have implications in reinterpreting the results of such models. There are also several studies which suggest that the depletion induced collapse mechanism is enhanced with crowders of sizes much smaller than monomers of the polymer~\cite{kang2015effects,zosel2020depletion,sung2021smaller,chen2019comparative,mardoum2018crowding,latshaw2015effects}. Our results show that for crowders with attractive interactions, a similar degree of compaction can be achieved with the crowder size same as that of monomers with tuneable  crowder-crowder interactions. 

In addition, our results can also have relevance for phenomena of cosolvency and cononsolvency exhibited by mixed solvents~\cite{masegosa1984preferential, tanaka2008temperature,mukherji2017depleted}. Mixtures of solvents can have significant and often counterintuitive effects  on the polymer conformation. For instance, a mixture of two good solvents can act as an effective poor solvent as in the phenomenon of cononsolvency. There are aspects about these counterintuitive phenomenon which are still controversial~\cite{bharadwaj2019does,mukherji2014polymer,mukherji2016relating,bharadwaj2021interplay}: whether a generic nonspecific attraction between polymer and cosolvent or preferential adsorption of cosolvent can explain the cononsolvency phenomenon. The problem of a polymer with attractive crowders can be recast  as one of a polymer with two solvents where one of the solvent in treated implicitly, via intra polymer attractive interaction, as considered in this paper. By varying $\epsilon_{mm}$, $\epsilon_{mc}$ and $\epsilon_{cc}$ independently along with varying $\rho_c$, some of these aspects can be explored. Understanding this mapping better is a promising area for future study.

\begin{acknowledgments}
	The simulations were carried out on the high performance computing machines Nandadevi at the Institute of Mathematical Sciences.
\end{acknowledgments}

\bibliography{neutral}
\newpage
\beginsupplement

\begin{center} \bf{Supporting Information}
\end{center}

\begin{figure}[h!]
	\includegraphics[width=0.6\columnwidth]{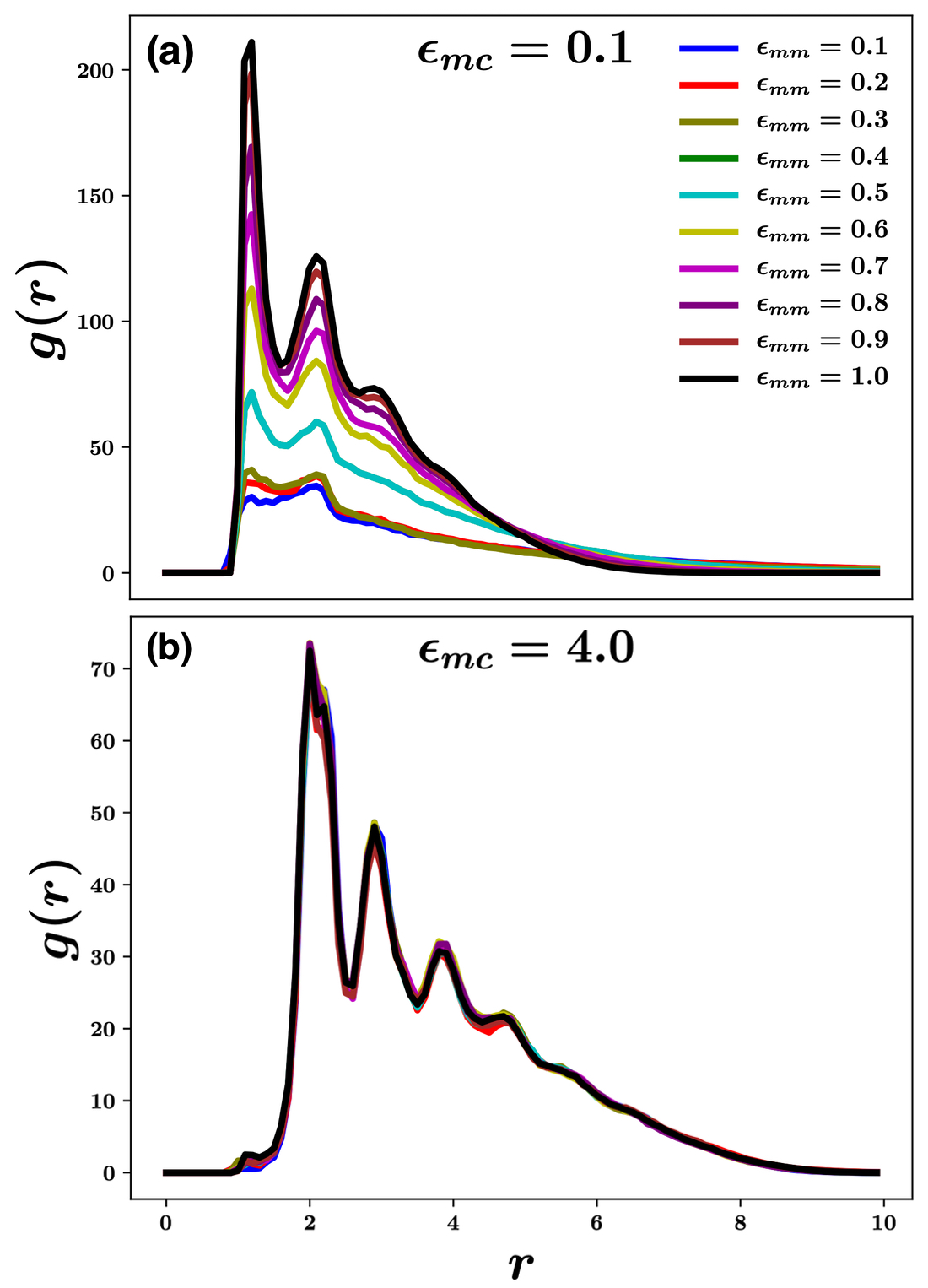}
	\caption{Monomer-monomer radial distribution function for (a) $\epsilon_{mc}=0.1$ and (b) $\epsilon_{mc}=4.0$ for different values of $\epsilon_{mm}$. The nearest bonded monomer contribution is excluded. The data are for  $\phi_c=0.047$ and $\epsilon_{cc}=1.0$. }
	\label{fig:grcompare}
\end{figure}

In Fig.~\ref{fig:grcompare}, the monomer-monomer radial distribution function is shown for $\epsilon_{mc}=0.1$ and $\epsilon_{mc}=4.0$ for different values of $\epsilon_{mm}$. In the $CI$ phase (Fig.~\ref{fig:grcompare}(a)), $g(r)$ is strongly dependent on $\epsilon_{mm}$ and has a sharp peak at $r\approx1.1$. In the $CB$ phase the first peak is absent and in addition $g_{mm}(r)$ is largely independent of $\epsilon_{mm}$.

\begin{figure}[h!]
	\includegraphics[width=0.6\columnwidth]{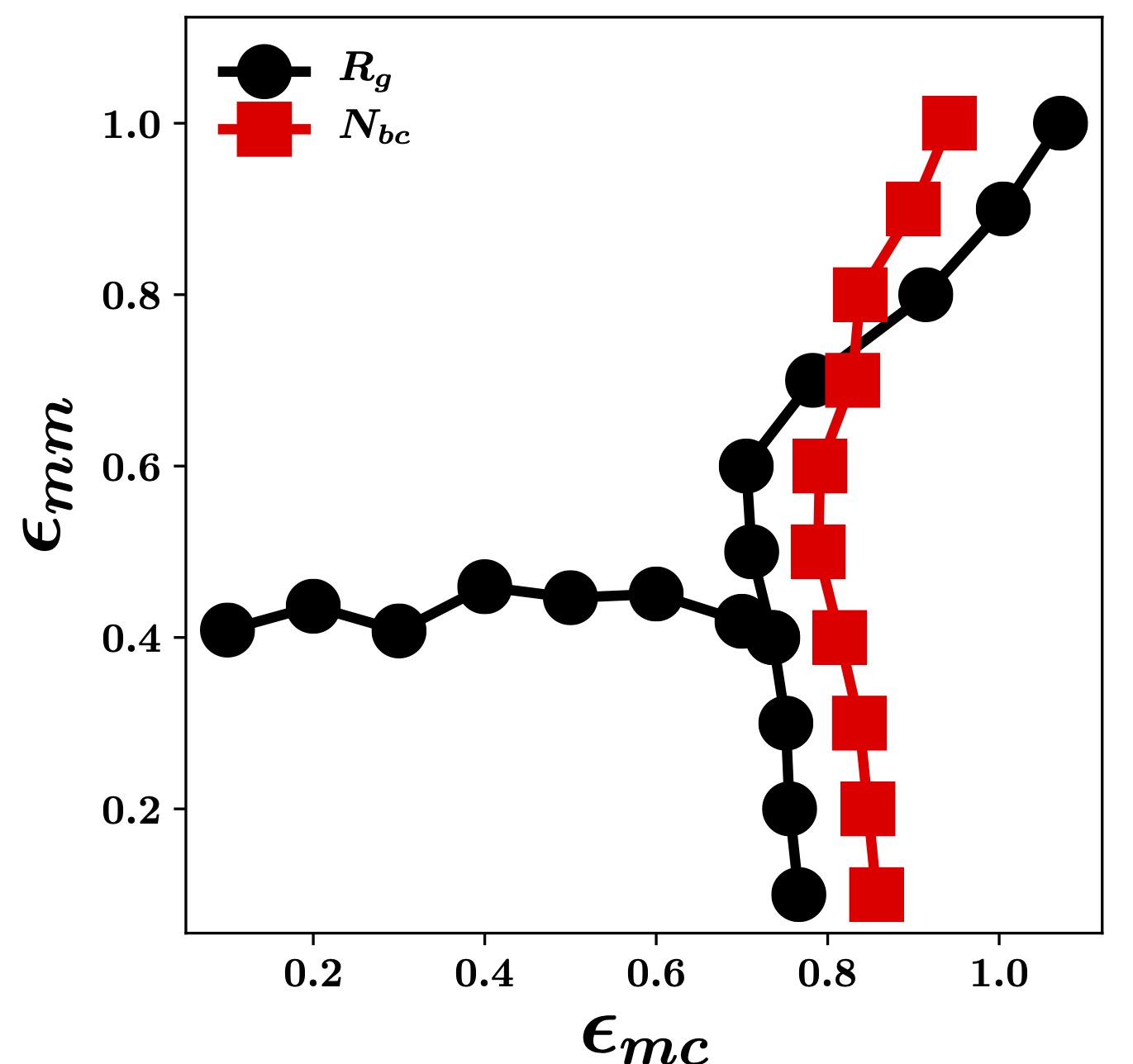}
	\caption{Comparison of the conformational phase diagram of the polymer in the $\epsilon_{mm}$-$\epsilon_{mc}$ plane obtained from analyses of $R_g$ and bridging crowder data. The data are for $\phi_c=0.047$ and $\epsilon_{cc}=1.0$.}
	\label{fig:comprgnbc}
\end{figure}

We calculate $\epsilon_{mc}^*$ from the $N_{bc}$ data following the same procedure as was done with $R_g$ data and the phase diagram, thus obtained, is shown in Fig.~\ref{fig:comprgnbc}.  We find the phase diagram obtained using $R_g$ and $N_{bc}$ are qualitatively the same, thus underscoring the robustness of the identification of the phases and the phase diagram in Fig.~2 (see main text)s.

\begin{figure}[h!]
	\includegraphics[width=0.6\columnwidth]{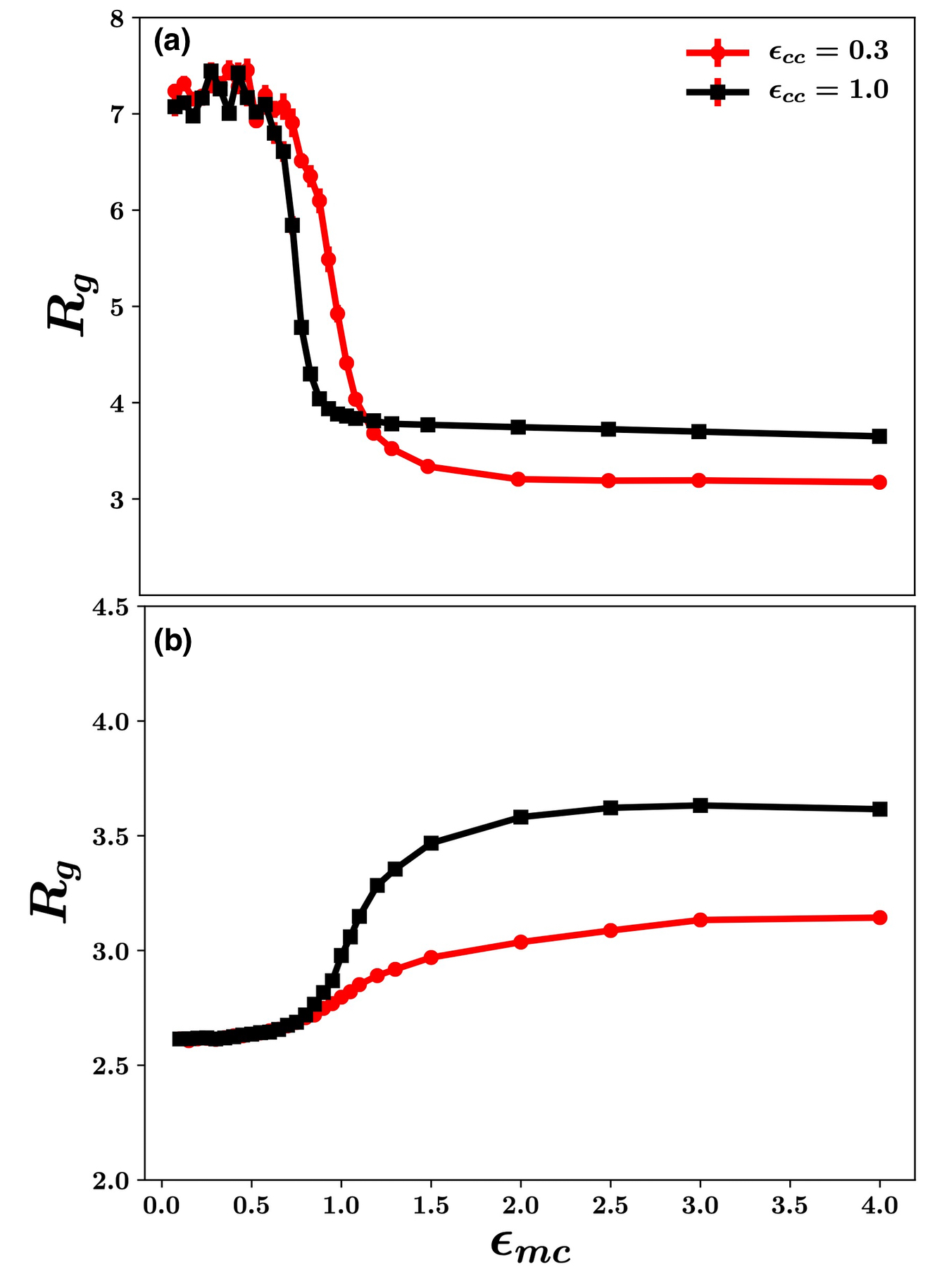}
	\caption{Comparison of the variation of $R_g$ with $\epsilon_{mc}$ for $\epsilon_{cc}=0.3,1.0$ when (a) $\epsilon_{mm}=0.1$ and (b) $\epsilon_{mm}=1.0$. The data are for $\phi_c=0.047$. }
	\label{fig:grcompare1}
\end{figure}

In Fig.~\ref{fig:grcompare1}, we show the effect of $\epsilon_{cc}$ on the $E$--$CB$ and $CI$--$CB$ transition lines. The $E$--$CB$ transition occurs at a larger value of $\epsilon_{mc}$ when $\epsilon_{cc}$ is decreased while the $CI$--$CB$ transition occurs at a smaller value of $\epsilon_{mc}$ when $\epsilon_{cc}$ is decreased. 

\begin{figure}[h!]
	\includegraphics[width=0.6\columnwidth]{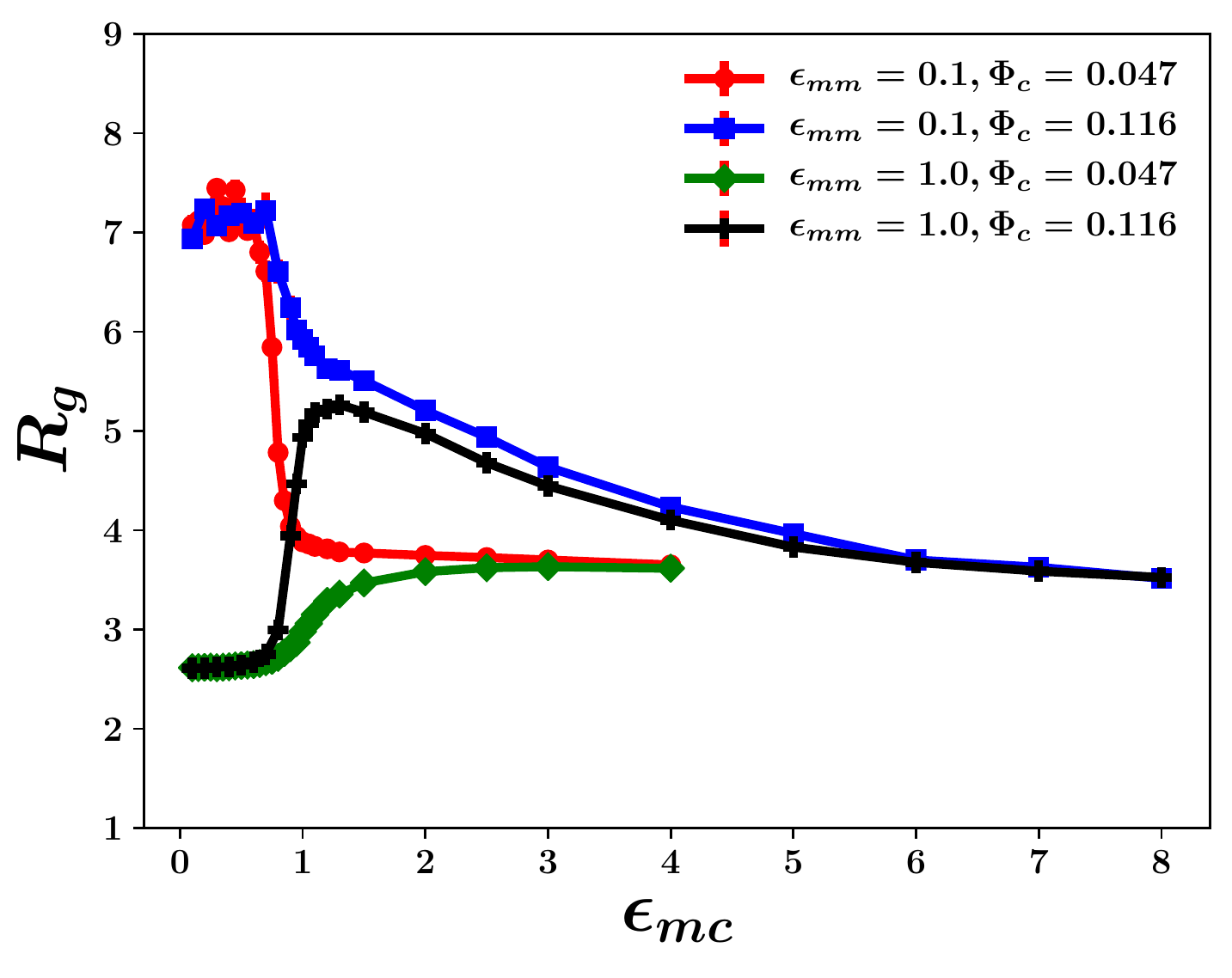}
	\caption{Comparison of the variation of $R_g$ with $\epsilon_{mc}$ for $\phi_c=0.047,0.116$ when $\epsilon_{mm}=0.1$ and $\epsilon_{mm}=1.0$. The data are for $\epsilon_{cc}=1.0$.}
	\label{fig:grcompare2}
\end{figure}

When crowder density $\phi_c$ is increased, there is a significant difference in the $R_g$ values in the $CB$ phase (see Fig.~\ref{fig:grcompare2}), likely due to increased number of bridging crowders present inside the collapsed phase. Only when $\epsilon_{mc}$ is significantly increased, the $R_g$ at $\phi_c=0.116$ approaches the limiting value of $R_g$ at $\phi_c=0.047$. 

\end{document}